\documentclass[prb,aps,multicol,amssymb,epsfig]{revtex4}
\usepackage{epsfig}

\begin{document}

\title{Exact Activation Energy of Magnetic Single Domain Particles}
\author{Daniel Braun\footnote{ Corresponding author. Fax: +1 914 945
4421; email: v2braun@us.ibm.com}}
\address{Infineon Technologies, 2070 State Route 52, Hopewell Junction, NY, 12533\\
MRAM Developement Alliance, IBM/Infineon Technologies, IBM Semiconductor
Research and Developement Center, 2070 State Route 52, Hopewell Junction,
NY, 12533}

\widetext
\begin{abstract}
\begin{center}
\parbox{14cm}{I present the exact analytical expression for the activation
energy as a 
function of externally 
applied magnetic fields for a single--domain magnetic particle with uniaxial
anisotropy (Stoner--Wohlfahrt model), and investigate the scaling behavior
of the activation energy close to the switching boundary. 
}
\end{center}
\end{abstract}
\maketitle

\section{Introduction}
A lot of effort has been spent over the last few years to understand the
magnetization reversal of small magnetic particles \cite{Lederman94,Kent94,Wernsdorfer95,Koch98,Bonet99,Wernsdorfer00,Schumacher03}.  
At sufficiently low temperatures, macroscopic quantum tunneling has been
observed \cite{Coppinger95,Friedmann96,Wernsdorfer97,Bokacheva00}, while for higher temperatures thermally activated behavior 
may switch the magnetization of the particle \cite{Koch00}. For the anlaysis of the 
experiments one needs to know the activation energy. Moreover, with the
advancing  
developement of integrated magnetoresistive memory devices (MRAM) the
dependence of the energy barrier on the magnetic field has become of
crucial technological importance as well. In a typical MRAM array, magnetic
memory cells are written by a coincident field technique, where both a
selected bitline (BL) and a selected wordline (WL) create magnetic fields,
the sum of 
which are strong enough to switch the memory cell, whereas the fields from
either BL or WL alone are
not sufficient to switch the cell. Nevertheless, these fields do
destabilize the non--selected cells to some extent, i.e. they reduce the energy
barrier against 
thermally activated switching. Also, even for the selected cells the
switching at 
finite temperatures happens before the actual zero temperature boundary of stability is 
reached, again due to thermal activation during the finite duration of a
write pulse \cite{Koch00,Braun93}. To estimate the life time of the information
in the memory, one needs to know the dependence of the energy barrier
on the applied fields as precisely as possible, as the energy barrier enters the switching rate exponentially. 

In the study of the switching behavior of small size magnetic
particles, the 
Stoner Wohlfarth model plays a central role \cite{Stoner48}. It describes
a single domain particle with uni--axial anisotropy in an external magnetic
field. The single domain approximation greatly simplifies the analysis, and
becomes a good approximation if the size of the particle becomes comparable
to or smaller than the exchange length, which in memory elements etched out of
a thin magnetic film is typically of the order of 100nm. The activation
energy in the Stoner Wohlfarth model can be 
calculated trivially 
if the external field is aligned either parallel to the preferred
axis or perpendicular to it. However, so far no analytical
solution has been known for the general case of the external field
pointing in an arbitrary direction \cite{Wernsdorfer03}. Given the crucial
importance of the field dependence of the activation energy, an exact
analytical solution will be provided in the present paper.

\section{The Stoner Wohlfarth Model}  
The energy of a uniformly magnetized particle with uniaxial symmetry,
characterized by the anistropy energy density $K$ and saturation
magnetization $M_s$  depends
on its magnetization via the angle $\vartheta$ between the magnetization
and the preferred axis,
\begin{equation} \label{Estowo}
E(\vartheta)=KV \sin^2\vartheta -V M_s H_x\cos\vartheta-V M_s H_y\sin\vartheta\,,
\end{equation}
where $H_x$ and $H_y$ are the magnetic field components  parallel and
perpendicular to the preferred axis in a plane containing the magnetization and the preferred axis, respectively, and $V$ denotes the 
volume of the sample \cite{Stoner48}. In the following
dimensionless variables will be used by writing the energy in terms of
$2KV$, $e(\vartheta)=E(\vartheta)/(2KV)$, and the magnetic field components in
terms of the switching field $H_c=2K/M_s$ as $h_x=H_x/H_c$, $h_y=H_y/H_c$,
such that
\begin{equation} \label{estowo}
e(\vartheta)=\frac{1}{2}\sin^2\vartheta -h_x\cos\vartheta-h_y\sin\vartheta\,.
\end{equation}   
For vanishing magnetic fields, this model  has a bistable ground
state, whereas for very large fields ($h_x\gg 1$ or $h_y\gg 1$) the first term can be neglected, and there is
only one minimum in the interval $-\pi\le\vartheta<\pi$, such that the magnetization
tends to align parallel to the applied field. Thus, for increasing field,
one of the original minima has to disappear, and the fields where this
happens mark the boundary of bistability. These fields are easily obtained
by setting both the first and second derivative of (\ref{estowo}) to zero
and eliminating $\vartheta$, whereupon the famous Stoner-Wohlfahrt astroid 
$h_x=\cos^3 \vartheta$, $h_y=\sin^3\vartheta$ is obtained \cite{Stoner48}. 
\section{Activation Energy}
The activation energy can in principle be calculated in a straight forward
manner by
finding the two minima and the two maxima of the energy in the bistability
range, determining the metastable of the two minima and the energy
barrier to the  smaller of the two maxima \cite{barrier}. In the case of
$h_x=0$ or $h_y=0$ this is straight forward: For $h_y=0$, 
\begin{equation} \label{d1e}
\frac{\partial e}{\partial
\vartheta}=\cos\vartheta\sin\vartheta+h_x\sin\vartheta-h_y\cos\vartheta=0
\end{equation}
leads to $\vartheta=0$, $\vartheta=\pi$, or $\cos\vartheta=-h_x$. The 
solution $\vartheta=0$ is metastable for $-1\le h_x\le 0$, stable for
$h_x>0$, and unstable for
$h_x<-1$. Correspondingly, $\vartheta=\pi$ is metastable for $0\le h_x<1$,
stable for $h_x<0$, and unstable for $h_x>1$. The solution
$cos\vartheta=-h_x$ leads
to two maxima which become complex for $|h_x|>1$. Thus,
the energy barrier for switching from $\vartheta=\pi$ to $\vartheta=0$ is
given by 
$E_A=e(\arccos(-h_x))-e(\pi)=(1-h_x)^2/2$. Similarly, for $h_x=0$ one finds
the energy barrier $E_A=(1-h_y)^2/2$. Therefore, if one of the two field
components vanishes, the activation energy depends quadratically on the
distance from the stability boundary. 

\begin{center}
\begin{figure}[h]
\epsfig{file=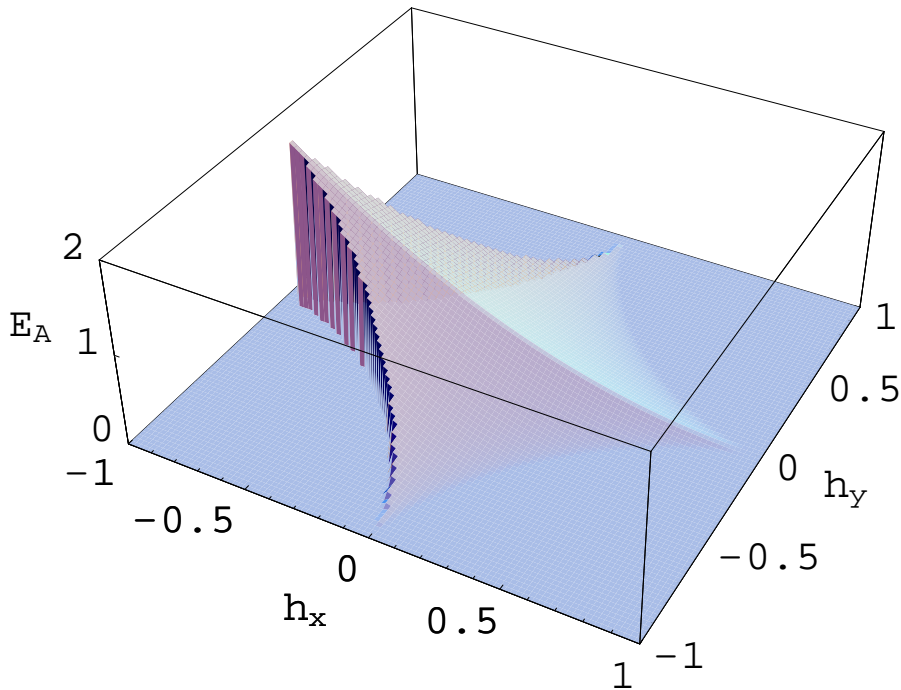,width=12cm,angle=0} \\[0.2cm]
\caption{Activation energy as function of $h_x$ and $h_y$, out of the state
stable for $h_x<0$. Outside the astroid there is only one stable and
no metastable
state, and the activation energy is therefore undefined.}\label{fig.EA3D}  
\end{figure}
\end{center}

In the general case, where neither $h_x$ nor $h_y$ vanish, one may
substitute $\sin\vartheta=u$, $\cos\vartheta=\pm\sqrt{1-u^2}$ into
(\ref{d1e}). Squaring the equation leads to
\begin{equation} \label{d1eu}
-u^4+2u^3h_y+u^2(1-h_x^2-h_y^2)-2uh_y+h_y^2=0\,.
\end{equation}
In order to calculate the energy barrier, one needs to find the roots of
this 4th order polynomial, which makes the analysis much more
cumbersome than for $h_xh_y=0$. Still, the roots of a 4th order
polynomial can be obtained analytically. The solutions $\vartheta_i$ are
conveniently written in terms of the functions 
\begin{eqnarray}
f_1&=&h_x^2+h_y^2-1\,,\\
f_2&=&108 h_x^2h_y^2+2f_1^3\,,\\
f_3&=&f_2+\sqrt{f_2^2-4f_1^6}\,,\\
f_4&=&\sqrt{1+\frac{f_1}{3}+\frac{2^{1/3}}{3}\frac{f_1^2}{f_3^{1/3}}+\frac{1}{3}\frac{f_3^{1/3}}{2^{1/3}}-h_y^2}\,.
\end{eqnarray}
Eq.(\ref{d1eu}) has four solutions for $u$, but the ambiguity in the sign of
$\cos\vartheta$ in terms of $u$ leads at this point to eight solutions for
$\vartheta$. They differ by three signs $\mu,\nu,\sigma$ in various places, and
are given by  
\begin{eqnarray}
\vartheta_i&=&\sigma \arccos\Bigg[-\frac{h_x}{2}+\frac{\mu}{2}f_4
+\frac{\nu}{2}\sqrt{1-\frac{f_1}{3}-\frac{2^{1/3}f_1^2}{3f_3^{1/3}}-\frac{1}{3}\frac{f_3^{1/3}}{2^{1/3}}+h_x^2-h_y^2+\frac{2\mu}{f_4}(2h_x+f_1h_x-h_x^3)}\Bigg]\,.\label{theti}
\end{eqnarray}
The signs will be chosen according to the binary decomposition
$(\mu,\nu,\sigma)=(-,-,-)$ for $\vartheta_1$, $(-,-,+)$ for $\vartheta_2$, $(-,+,-)$ for
$\vartheta_3$, $\ldots$, $(+,+,+)$ for $\vartheta_8$. None of these
solutions solves 
(\ref{d1e}) for all $h_x,h_y$. Rather, each of them solves the equation only
in two quadrants. However, it is possible to construct four uniform
solutions out of the eight partially valid ones, which are continuous (after
the identification of $\pi$ with $-\pi$) and solve
(\ref{d1e}) for all values of $h_x,h_y$. The four uniform solutions are  
\begin{eqnarray}
\tilde{\vartheta}_1&=&\theta(h_y)\vartheta_2+\theta(-h_y)\vartheta_1\,,\\
\tilde{\vartheta}_2&=&\theta(h_xh_y)\vartheta_5+\theta(-h_xhy)\vartheta_3\,,\\
\tilde{\vartheta}_3&=&\theta(h_xh_y)\vartheta_4+\theta(-h_xhy)\vartheta_6\,,\\
\tilde{\vartheta}_4&=&\theta(h_y)\vartheta_8+\theta(-h_y)\vartheta_7\,,\\
\end{eqnarray}
where $\theta(x)=1$ for $x>0$ and zero elsewhere denotes the Heaviside
function. 
Calculating the second derivative one finds that the first uniform solution
is a minimum for $h_x<0$. It disappears as real solution for $h_x>0$ outside
the astroid. Thus, this is the metastable state for $h_x>0$. On the other
hand, $\tilde{\vartheta}_4$ remains a minimum even outside the astroid for
$h_x>0$, but disappears as real solution outside the astroid for
$h_x<0$. The uniform solutions $\tilde{\vartheta}_2$ and
$\tilde{\vartheta}_3$ are both maxima inside the astroid, and become complex
outside one half of the astroid (for $h_y<0$ and $h_y>0$, respectively {\bf
??}).  One easily convinces oneself that for $h_y\ge 0$, $\tilde{\vartheta}_3$
is the relevant maximum ($e(\tilde{\vartheta}_3)\le e(\tilde{\vartheta}_2)$),
whereas for $h_y<0$, escape over the maximum at $\tilde{\vartheta}_2$ is
dominant ($e(\tilde{\vartheta}_3)>e(\tilde{\vartheta}_2)$). 
The activation energy out of the metastable state is therefore given by 
\begin{eqnarray} 
E_A&=&e(\tilde{\vartheta}_3)-e(\tilde{\vartheta}_1)\mbox{ for }
h_y\ge 0\,,\label{eAhyg0}\\
E_A&=&e(\tilde{\vartheta}_2)-e(\tilde{\vartheta}_1)\mbox{ for }
h_y<0\,,\label{eAhyl0}
\end{eqnarray}
Fig.\ref{fig.EA3D} shows a plot of the activation energy as function of
$h_x$ and $h_y$. 

\noindent
\vspace*{0.1cm}

For $h_y=0$ both maxima lead to the same activation energy, and
$E_A$ is an even function of $h_y$, as it should be. In the following
the attention will therefore focus solely on the case $h_y\ge 0$, and to the
case where the initial state $\vartheta=\pi$ is metastable, $h_x>0$,
i.e.~the first 
quadrant. In principle, the particle might 
be excited also out of the stable state and end up (for a finite time) in
the metastable state, but this process is of much less importance, as the
activation energy is much larger and it enters exponentially in the thermal
switching rate.  

\section{Scaling}
\begin{center}
\begin{figure}[h]
\epsfig{file=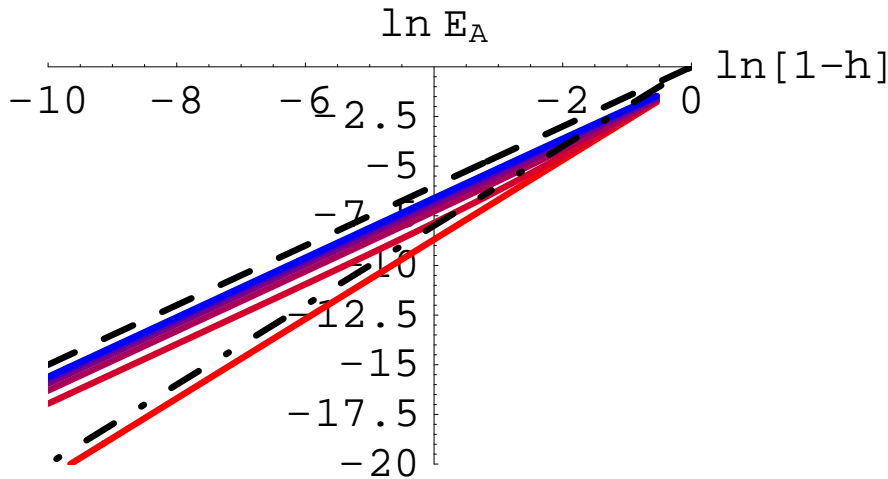,width=12cm,angle=0} \\[0.2cm]
\caption{Scaling of the activation energy $E_A$ as function of 
the relative distance $h$ from the astroid boundary ($h_x=h
\cos^3\xi$, $h_y=h \sin^3\xi$). The curves show
decreasing slopes for values of $x=\xi/\pi=0$ (red), $1/24$, (pink)
$2/24,\ldots 1/4$ (blue). The dashed line corresponds to a power law with
exponent 3/2, the dot--dashed line to an exonent 2.}\label{fig.EAloglog}  
\end{figure}
\end{center}

The scaling of the activation energy as function of the distance $1-h$ from the
astroid boundary is of  particular interest. Here, $h$ is defined by $h_x=h
\cos^3\xi$, $h_y=h \sin^3\xi$ such that $h=1$ always corresponds to the
astroid boundary. It has been shown before \cite{Wernsdorfer96} that
the scaling must be of the form  
\begin{eqnarray}
E_A&=&c_{3/2}(\xi)(1-h)^{3/2}+c_{2}(\xi)(1-h)^{2}+c_{5/2}(\xi)(1-h)^{5/2}+\ldots\,.\label{E_As}
\end{eqnarray}
All coefficients besides $c_{2}(\xi)$ vanish for $\xi=0$, and  but it has
been shown 
numerically \cite{Wernsdorfer96} and theoretically \cite{Victora89} that
already for small values of $\xi$ the coefficient $c_{3/2}$ dominates the
scaling. This is confirmed and made more precise by using the exact
solution.  
Fig. \ref{fig.EAloglog} shows
$\ln E_A$ as function of $\ln(1-h)$. The plot reveals power law behavior to
a good 
approximation even rather far away from the astroid boundary, $E_A\propto
(1-h)^a$  with an exponent $a=2$ for $\xi=0$, and 
an exponent close to $3/2$ for larger values of $\xi$.  

\noindent
\vspace*{0.1cm}

\begin{center}
\begin{figure}[h]
\epsfig{file=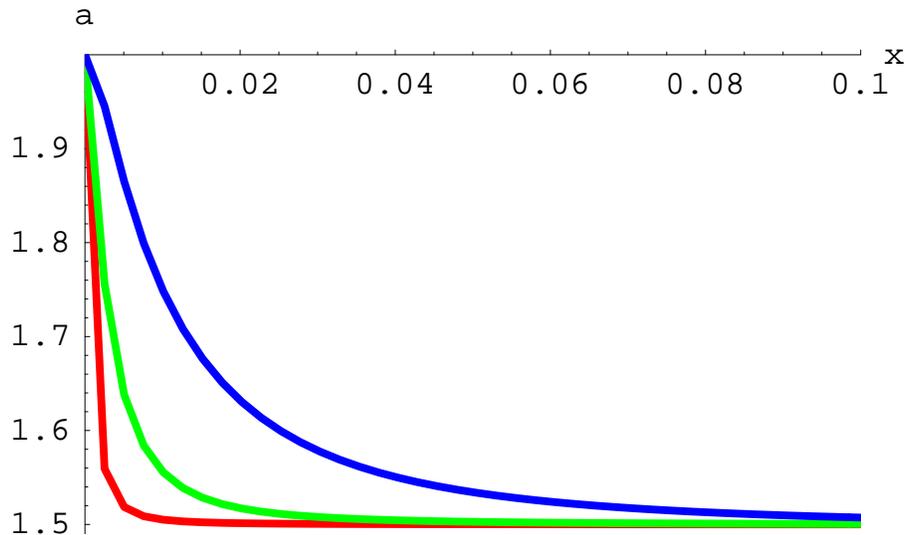,width=12cm,angle=0} \\[0.2cm]
\caption{Scaling exponent $a$ extracted from a fit of $\ln E_A$ to
$a\ln(1-h)$, using the exact expression for $E_A$, eq.(\ref{eAhyg0}). The
three curves correspond to fitting ranges $h=0.93187079$ to $h=0.996837$
(top curve --- blue), $h=0.99538$ to $h=0.999978$ (middle curve --- green),
and $h=0.9996837$ to $h=0.999998532$ (bottom curve --- red). For all
calculations of the activation energy, 80 digits precision was
used.}\label{fig.alin}   
\end{figure}
\end{center}

According to the numerical evaluation of (\ref{eAhyg0}) the
exponent is symmetric with respect to $\xi=\pi/4$. 
The exact value of the exponent depends on the fitting range and
on whether a quadratic term is included in the fit of $\ln E_A$ as
function of $\ln(1-h)$. Fig.~\ref{fig.alin} shows the fitted exponent $a$
as a function of $x=\xi/\pi$ for three different 
fitting  ranges, from $h=0.932$ to $h=0.99999853$, assuming a pure power
law, $\ln E_A=a\ln(1-h)$. The observed dependence of the exponent on $x$
is very similar to what was previously calculated numerically
\cite{Wernsdorfer96}. In particular, the exponent appears to become even
slightly smaller than $3/2$ for values of $\xi$ close to $\pi/4$. 

\noindent
\vspace*{0.1cm}

\begin{center}
\begin{figure}[h]
\epsfig{file=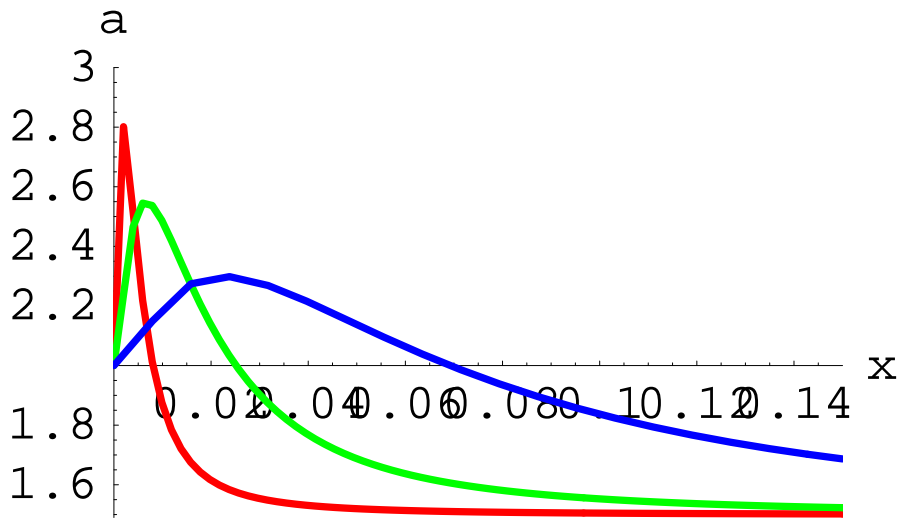,width=12cm,angle=0} \\[0.2cm]
\caption{Scaling exponent $a$ extracted from a fit of $\ln E_A$ to
$a\ln(1-h)+b\ln^2(1-h)$. The 
three curves correspond to the same fitting ranges used in
Fig.\ref{fig.alin}. }\label{fig.aquad}   
\end{figure}
\end{center}

However, there is a substantial non--linear part in the scaling behavior, as
becomes obvious when fitting to 
$\ln E_A=a \ln(1-h) +b \ln^2(1-h)$. The extracted linear part is plotted
in Fig.\ref{fig.aquad}. For small values of
$\xi$, the exponent can now be substantially larger than $2$. 

\noindent
\vspace*{0.1cm}

\section{Summary} I have derived the exact analytical expression of the
activation energy for single domain switching of small magnetic particles in arbitrary magnetic fields
(Stoner--Wohlfarth model). The activation energy scales
approximately like a power law as a function of the distance of the
switching boundary (astroid) up to distances of order unity, but also
contains a substantial non--power law term. 

{\em Acknowledgement:} I am grateful to Daniel Worledge for a useful
discussion.



\begin{references}
\bibitem{barrier} Note that in principle there are two energy barriers,
corresponding to the two escape paths out of the metastable minimum by
clockwise or counterclockwise rotation of the magnetization. In the
following the path with the higher activation energy will be neglected and
the activation energy will always be defined as the smaller of the two
energy barriers.
\end{references}
\end{document}